\documentclass[12pt,preprint]{aastex}
\slugcomment{IPM/P-2003/067} \shorttitle{Possibility of Magnetic
Mass detection by the next generation microlensing experiments. }
\shortauthors{Sohrab Rahvar \& Farhang Habibi }
\begin{document}

\title{ Possibility of Magnetic Mass Detection by the Next Generation of Microlensing Experiments}


\author{Sohrab Rahvar \altaffilmark{1,2} and Farhang Habibi \altaffilmark{3}}
\email{rahvar@sharif.edu}


\altaffiltext{1}{Department of Physics, Sharif University of Technology,
 P.O.Box 11365--9161, Tehran, Iran}
\altaffiltext{2}{Institute for Studies in Theoretical Physics and
Mathematics, P.O.Box 19395--5531, Tehran, Iran}
\altaffiltext{3}{Department of Physics, Iran University of Science
and Technology, Narmak, Tehran 16844, Iran}
\begin{abstract}
We study the possibility of magnetic mass detection by the
gravitational microlensing technique. Recently the theoretical
effect of magnetic mass in the NUT space on the microlensing light
curve has been studied. It was shown that in the low photometric
signal to noise and sampling rate of MACHO experiment light
curves, no signature of NUT factor has been found. In order to
increase the sensitivity of magnetic mass detection, we propose a
systematic search for microlensing events, using the currently
running alert systems and complementary telescopes for monitoring
the Large Magellanic Clouds stars. In this strategy of
observation, we obtain the magnetic mass detection efficiency and
also the lowest observable limit of the NUT factor. This method of
survey for gravitational microlensing detection can also be used
as a tool for searching other exotic space-times.
\end{abstract}
\keywords{gravitational lensing -- relativity -- cosmology:
Observations -- Cosmology: theory --dark matter.}
\section{Introduction}
The gravitational microlensing method for detecting MAssive
Compact Halo Objects (MACHOs) in the Milky Way halo has been
proposed by Paczy\'nski (1986). Many groups have contributed to
this experiment and have detected hundreds of microlensing
candidates in the direction of the galactic bulge, spiral arms and
Large and Small Magellanic Clouds (LMC \& SMC). Due to the low
probability of the microlensing detection, less than 20 events
have been observed by the EROS and MACHO groups in the direction
of the Magellanic clouds (Lasserre et al. 2000; Alcock et al.
2000). The low statistics not only causes ambiguities in
identifying the galactic model of the Milky Way, but also in some
cases the microlensing results
are at variance with the results of other observations (Gates and Gyuk 2001).\\
Comparing LMC microlensing events with the theoretical galactic
models can give us the mean mass of MACHOs and the fraction of
halo mass in the form of MACHOs. In the case that we use a Dirac
Delta mass function for the MACHOs, the mass of MACHOs in standard
halo model is obtained about $\sim 0.5 M_{\sun}$. This means that
the initial mass function of MACHO progenitors in the galactic
halo should be different from that of the disk, because we neither
see the low mass stars which should still exist nor heavier stars
that would have exploded in the form of supernova (Adams \&
Laughlin 1996; Chabrier, Segretain \& Mera 1996). Another
contradiction is that if there were as many white dwarfs in the
halo, as suggested by the microlensing experiments, they would
increase the abundance of heavy metals via Type I Supernova
explosions (Canal., Isern $\&$ Ruiz-Lapuente 1997). Also, recently
Green \& Jedamzik (2002) and Rahvar (2004) showed that the
observed distribution of duration of microlensing events is not
compatible with what is expected from the standard and
non-standard halo models. They showed that the observed
distribution is significantly narrower
compared to what is expected from the galactic models.\\
The mentioned problems can be a motivation for establishing the
next generation of the microlensing experiments. The new surveys
will have the potential to increase the number of microlensing
candidates and reduce the ambiguities due to Poisson statistics.
The other improvements of the new surveys can be the higher
sampling rates and the higher precision photometry of the light
curves. More precise light curves will enable us to distinguish
the deviations between the standard and non-standard light curves
due to parallax or source finite-size effects (Rahvar et al.
2003). In the so-called non-standard microlensing candidates the
degeneracy partially can be broken between the lens parameters,
such as the distance and the mass of a lens. A better
determination of the distance and the mass distributions of the
lenses can help us to better identify the Milky
Way halo model (Evans 1994).\\
Although the mentioned effects are in the background of a
Schwarzschild space, it is also be possible that a MACHO which
plays the role of the lens, resides in an exotic space-time such
as the Kerr or the NUT space. Deviation of the space-time from the
Schwarzschild metric causes deviation of the microlensing light
curve from the standard one. Thus, studying the microlensing light
curves not only can be used to determine the dark matter in the
form of MACHOs but also as a unique tool to explore the other
exotic space-times as well.\\
In the paper by Nouri-Zonoz and Lynden-Bell (1997) the
gravitational lensing effect on the light rays passing by a NUT
hole has been considered, using the fact that all the geodesics in
the NUT space, including the null ones, lie on cones. The
extension of this work to the microlensing light curve in the NUT
space has been studied by Rahvar and Nouri-Zonoz (2003) and the
possible existence of magnetic mass on the light curves of the
MACHO group microlensing candidates has been tested. According to
the analysis of the light curves, no magnetic mass effect has been
found. Although the result showed that the effect of the NUT
factor is almost negligible, one can not rule out the existence of
NUT charge on that basis. The next generation of microlensing
experiments may prove the (non-) existence of magnetic mass
through a more careful study of the microlensing light curves.\\
Here in this work we simulate the microlensing light curves in the
NUT metric according to a strategy for the next generation of
microlensing surveys. The aim of this work is to obtain the
observational efficiency for the magnetic mass detection and to
find the lowest limit for the NUT charge that can be observed.
Large Magellanic Cloud (LMC) stars are chosen as the target stars
for monitoring. The advantage of using LMC stars as compared to
the spiral arms and the galactic bulge stars is the lower
contamination by blending and source finite-size effects, which
can affect the NUT light curves. The other advantage of LMC
monitoring is that it enables us to increase the microlensing
statistics to put a better limit on the mass of
the lenses and the mass fraction of the galactic halo in form of MACHOs. \\
The outline of this paper is as follows. In Section 2, we give a
brief account on the microlensing light curve in the NUT metric
and compare it with the Schwarzschild case. In Section 3, we
introduce the observational strategy and perform a Monte-Carlo
simulation to generate the microlensing light curves. Section 4
contains the fitting process to the simulated light curves to
obtain the observational efficiency of the magnetic mass
detection. The results are discussed in Section 5.
\section{Gravitational microlensing in Schwarzschild and NUT metrics}
The gravitational lensing effect occurs when the impact parameter
of a lens with respect to the un-deflected observer-source line of
sight is small enough that the deviation of source shape becomes
detectable. In the case that of a point like source, the
deflection angle is too small to be resolved by the present
telescopes. This type of gravitational lensing which amplifies the
brightness of the background star is called the gravitational
microlensing. In the Schwarzschild metric the magnification is
given by (Paczy\'nski 1986):
\begin{equation}
\label{pac}
A(t) = \frac{u(t)^2 +2}{u(t)\sqrt{u(t)^2 + 4}},
\end{equation}
where $u(t) = \sqrt{u_0^2 + (\frac{t - t_0}{t_E})^2}$ is the
impact parameter (position of the source in deflector plane
normalized by the Einstein radius, $R_E$) and in which $t_E$ is
the Einstein crossing time (duration of event) defined by $t_E =
R_E/v_t$, where $v_t$ is the transverse velocity of deflector with
respect to the line of sight. The Einstein radius is given by $
R_E^2 = \frac{4GMD}{c^2}$, where M is the mass of the deflector
and $D = \frac{D_{l}D_{ls}}{D_{s}}$. $D_l$, $D_ls$ and $D_s$ are
the observer-lens, lens-source and observer-source distances,
respectively. The only physical parameter that can be obtained
from a light curve is the duration of the event which is a
function of the lens parameters such as mass, the distance of lens
from the observer and the relative transverse velocity of
the lens with respect to our line of sight.\\
In the case of gravitational microlensing, the configuration of
the lens changes within the time scales of dozen of days while in
the cosmological scales the lensing configuration is almost
static. Since the magnification factor depends on the space-time
metric, the gravitational microlensing technique may also be a
useful tool to explore the other exotic metrics like the NUT
space. In the NUT space the magnification due to the microlensing
depends in addition, to an extra factor (magnetic mass) compared
to the Schwarzschild space. It should be mentioned that the NUT
space reduces to the Schwarzschild one when the magnetic mass
'${\it l}$' is zero\footnote{NUT space is give by the following
metric: $ds^2 = f(r)(dt-2l\cos\theta d\phi)^2 -\frac{1}{f(r)}dr^2
-(r^2 +l^2)(d\theta^2 +sin^2\theta d\phi^2)$, where $f(r) = 1 -
2(Mr +l^2)/(r^2 + l^2)$.}. So we expect that the microlensing
amplification reduces to Equation (\ref{pac}) for zero magnetic
mass. Rahvar and Nouri-Zonoz (2003) obtained the magnification in
this space-time as follows:
\begin{equation}
\label{nut_amplification}
A(u) = \frac{2 + u^2}{u\sqrt{4 + u^2}} +
\frac{8R^4(2+u^2)}{u^3(4+u^2)^{3/2}} +{\mathcal{O}}(R^8)+ ...,
\end{equation}
where $R_{NUT}=\sqrt{2lD}$ is defined as the NUT radius (analogous
to the Einstein radius) and $l$ is the magnetic mass of the lens.
Parameter $R$ in Equation (\ref{nut_amplification}) is defined by
dividing NUT radius to the Einstein radius, $R =R_{NUT}/R_E$. It
is seen that in the NUT space the magnification factor, like the
in the Schwarzschild case, is symmetric with respect to time. The
extra second term implies a bigger relative maximum of the
magnification factor for a given minimum impact parameter. we have
also a shape deviation
of the light curve with respect to the case of Schwarzschild metric.\\
The detectability of the NUT factor through studying microlensing,
depends on the light curves quality (i.e. sampling rate and
photometric error bars). In the next section we introduce a new
strategy for microlensing observations in order to improve the
microlensing light curves, both from the point view of the
sampling rate and the photometric precision.
\section{Light curves simulation in the NUT space}
The strategy of the observation is based on using a survey as an
alert system for microlensing detection with a follow-up setup.
EROS is one of the groups that used an alert system to trigger
ongoing microlensing events. We simulate EROS like telescope with
the same sampling rate, considering $70\%$ clear sky at {\it La
Silla} during the observable seasons of the LMC. A follow-up
telescope is considered to observe with one percent photometry
precision and sampling rate of at least once per night those
events, that have been triggered by the first telescope. Here our
aim is to simulate microlensing light curves in the NUT space by
using the observational strategy, mentioned above.\\
It should be noted that there are at least two other important
effects which are called blending and source finite-size effects
that can change the light curves in symmetric manner like the NUT
factor. Those effects are important because they may dominant over
the effect of the NUT factor in the light curves. So, before
starting the simulation procedure we give a brief account on those
effects and include them in generating microlensing light curves
in the NUT
space.\\
The blending effect is due to the mixing of a lensed star and its
neighbors lights, which is given as:
\begin{equation}
F(t) = F_b + A(t)F_s,
\end{equation}
where $F(t)$ is the measured flux, $F_s$ is the lensed source,
$F_b$ is from the vicinity of lensed source and $A(t)$ is the
amplification (Wozniak and Paczynski 1997). This effect is
described by the blending parameter which is defined as $ b =
\frac{F_s}{F_s + F_b}$ and the observed magnification factor can
be written as
\begin{equation}
A_{obs}(t) = 1 + b(A(t) -1).
\end{equation}
The second altering effect on a light curve in a NUT space is the
source finite-size effect which is caused by the non-zero size of
projected source star on the lens plane. In this case, different
parts of the source star are amplified by different factors. The
relevant parameter of this effect is the projected size of the
source radius on the lens plane, normalized to the corresponding
Einstein radius ($U =\frac{xR}{R_E}$), where $x=\frac{D_l}{D_s}$
is the ratio of lens and source distances from the observer and
$R$ is the size of the source radius. In the case of close
source-lens distance compared to the observer-source
distance, this effect becomes important.\\
To find the best field of source stars, we compare possible fields
of observation such as the galactic bulge, the spiral arms and the
Magellanic clouds to find the least blending and source
finite-size effects. In the direction of galactic bulge the
blending effect is high, since the field of target stars is
crowded, except for the clump giants (Popowski et al 2000). For
the spiral arms stars the blending effect is less than towards the
galactic bulge while the source finite-size effect due to the
self-lensing by the spiral arms stars is considerable. For the
Small Magellanic Cloud (SMC) according to the blending and
parallax studies of long duration event (Palanque-Delabrouille et
al 1998), it seems that SMC is quite elongated along our line of
sight, with a depth varying from a few kpc (the tidal radius of
the SMC is of the order of $4$ kpc) to as much as $20$ kpc. So it
is seen that due to high blending and source finite-size effects,
SMC is not suitable for searching gravito-magnetic parameters. It
seems that LMC is the best choice for this study. The other
advantage of using this field is increasing the microlensing
statistics which
can be used in dark matter studies of the galactic halo.\\
Here in our simulation, we use the distribution of the blending
factor according to the reconstructed blending parameter that has
been obtained by the best fit to the LMC microlensing events. For
the source finite-size effects of LMC stars, which become
important in the case of itself-lensing, first we compare relative
self-lensing abundance as compared to the galactic halo lensing
and then evaluate the finite-size effect of
those events on the light curves. \\
Comparing the optical depth for the standard galactic halo model
$\tau_{halo}= 1.2^{+0.4}_{-0.3} \times 10^{-7}$ (Alcock et al.
2000) with the optical depth obtained by the LMC itself
$\tau_{self-lensing}=[0.47-7.84]\times 10^{-8}$ (Gyuk, Dalal and
Griest 2000), with the mean value of $2.4 \times 10^{-8}$ shows
that the expected microlensing events lensed by the halo MACHOs
are about one order of magnitude more than the LMC's. The optical
depth value of LMC self-lensing can be confirmed by studying the
parallax effect on the light curves. Rahvar et al. (2003) showed
that for using the same observational strategy that is proposed
here, if the self-lensing is dominant, very few lenses (only those
which belong to the disk) will produce
a detectable parallax effect. \\
In order to evaluate the source finite-size effect on the
microlensing light curves of LMC we perform a Monte-Carlo
simulation to produce the distribution of the relevant parameter
$U$. We use the LMC model introduced by Gyuk, Dalal and Griest
(2000) to see the matter distribution in our line of sight. The
probability of a microlensing event by a lens at LMC at a given
distance from us is
$$\frac{d\Gamma(x)}{dx}\propto \sqrt{x(1-x)}\rho(x), $$
where $\rho(x)$ is the matter density distribution of LMC. \\
The source stars at LMC are chosen according to their
color-magnitude distribution. We use the mass-radius relation
(Demircan and Kahraman 1990) to evaluate the radius of stars in
our simulation. The radius of source stars are projected on the
lens plane and normalized to the corresponding Einstein radius to
obtain the distribution of $U$s for the LMC self-lensing events.
The mean value of $U$ according to our simulation is about
$10^{-3}$, which we applied to obtain the gravitational
microlensing light curves. For an impact parameter as small as
$u_0=0.01$, where the NUT factor becomes important, the maximum
magnification difference of a standard light curve and that of
obtained by considering source finite-size effect is about one
percent. On one hand this difference is less than our photometric
accuracy and on the other hand the optical depth due to
self-lensing is one order of magnitude smaller as compared to the
galactic halo. The conclusion is that the source finite-size
effect is not important in our analysis.
\subsection{Simulation of light curves}
The aim of this section is to simulate the microlensing light
curves according to the observational strategy that was described
before. We use the theoretical light curves to fit the simulated
ones and evaluate the magnetic mass parameter of the NUT metric.
The final result in this procedure is the observational magnetic
mass detection efficiency, which can be applied in different
galactic models. To start simulating the light curves,
we use a uniform random function to generate the lens parameters.\\
The standard microlensing light curve in the Schwarzschild metric
depends on 4 parameters, namely the base flux, $u_0$ (minimum
impact parameter), $t_e$ (duration of the event) and $t_0$ (the
moment of minimum impact parameter or maximum magnification).
Taking into account the magnetic mass needs an extra parameter,
$R$. The relevant parameters in simulating the light curves are
chosen in the following intervals: $u_0\in[0,1]$, $t_0\in[0,2yr],$$t_E\in[5,365]$
days and $R\in[0,0.5]$.\\
The base fluxes $F_b$ of the background stars in the direction of
LMC are chosen according to the magnitude distribution in the EROS
catalogs (Lasserre 2000). Since it was shown that the contribution
of the blending effect is important in this study, we use the
blending distribution that has been obtained from the observed LMC
microlensing events in order to use them in generating the light
curves (Alcock et al. 2000). The light curves are simulated by
using the sampling rate of EROS which is about one observation per
six nights in average and is variable during the seasons. The
average relative photometric precision $\Delta F/F$ for a given
flux $F$ (in ADU unit) is taken from the EROS phenomenological
parametrization which has been found for a standard quality image
\cite{fre99}.
In simulating the light curves, every photometric measurement is
randomly shifted according to a Gaussian distribution that
reflects the photometric uncertainties. Since the photometric
uncertainty depends on the apparent magnitude of the background
stars, the error bars of light curves decrease by increasing the
brightness of background source during the lensing, (see Fig.
\ref{lc}).
\subsection{Simulation of a simple alert system}
The next step is to simulate an alert system to trigger the
ongoing events and the follow-up observation by the secondary
telescope.
According to one of the EROS alert algorithms, the events will be
announced as soon as their light curves exhibit 4 consecutive flux
measurements above 4 standard deviations from the base line
\cite{man}. It is clear that only the most significant
microlensing events are selected by this algorithm. We have in
fact considered several trigger thresholds, from a loose criterion
(3 consecutive measurements above $3\sigma$ from the base line) to
the strict criterion that was finally used. Even using this strict
criterion, in average one false alarm due to variable stars or
instrumental artifacts is expected per true microlensing alert
\cite{JFGprivate}. This false alarm rate will induce some lost
follow-up time, but for very limited durations, as it is usually
very fast to discard a non-microlensing event.
Fig. \ref{lc} shows an example of microlensing light curve that
has been simulated, using the specifications of the primary and
the secondary follow-up telescopes.\\
The efficiency of the alert system depends on the parameters of
the lenses. In order to obtain the trigger efficiency in terms of
the physical parameters such as the duration of events and $R$, we
integrate over the irrelevant parameters such as the minimum
impact parameter and the time of maximum magnification.  Equation
(\ref{nut_amplification}) shows that NUT parameter increases the
maximum magnification or in another word decreases the effective
minimum impact factor. The result is more trigger rate of
microlensing events for those that have larger $R$. This effect is
shown in Fig. \ref{trig_proj}. It shows that the trigger
efficiency is increased by the long duration of microlensing
events, which reflects a bigger probability for the observation of
long duration events as compared to short events.
\section{ Follow-up telescope and fitting process to the light curves}
We use a Monte-Carlo simulation to generate a large number of
microlensing events. At the first step the lens parameters are
chosen and the light curve is generated according to the primary
telescope specification. Using the trigger system, in the case
that an event is alerted, the secondary telescope starts its
measurements with high sampling rate and photometry precision of the
ongoing microlensing event.\\
The second telescope is supposed to be a partially dedicated
telescope which follows the measurements of alerted events. The
telescope is assumed to have about one percent precision in
photometry and perform the sampling of events through all the
clear nights. According to the Meteorological statistics of the
{\it La Silla} observatory about $70$ percent of nights per year
are clear. A one-meter telescope could achieve this
precision with a long exposure of about $30$ {\it min}. \\
After simulating a large number of events by this strategy, we use
the NUT and Schwarzschild theoretical microlensing light curves to
fit the simulated ones. The least square method is used to fit the
theoretical light curves on the data. An example of the fitting
routine is shown in Fig. {\ref{lc}. In the case of fitting data
with the NUT curve, with $R<0.1$ we encounter the degeneracy
problem of fitting. It means that for $R$ close to zero we may
obtain from the fitting a non-zero reconstructed value for $R$. To
distinguish between the microlensing light curves affected by NUT
charge and the standard ones we use the following criterion
denoted by $\Delta\chi^2$, to be more than two
\begin{equation}
\Delta\chi^2 = \frac{\chi_{Sch}^2 -
\chi_{NUT}^2}{\chi_{NUT}^2/N_{d.o.f}}\frac{1}{\sqrt{2N_{d.o.f}}},
\end{equation}
where indices of the $\chi^2$ correspond to the type of the metric
and $N_{d.o.f}$ is the number of degree of freedom in the NUT
fitting. As complementary criterion in addition to the mentioned
one we use the signal to noise ratio of $R$ to be more than two.
We obtain the magnetic mass detection efficiency of MACHOs by
dividing the reconstructed parameters of those events that pass
two mentioned criteria, to the generated events. Fig. \ref{2d_eff}
shows two dimensional efficiency of magnetic mass detection in
terms of $R$ and the duration of events.
The detection efficiency of magnetic mass has direct correlation
with $R$ as well as to the duration of the events. It is more
practical to have efficiencies in terms of duration of events and
$R$ which are shown in Fig.\ref{nut_eff}. It should be mentioned
that the blending effect decreases the detection efficiency of
magnetic mass, as shown in Fig.\ref{nut_eff}.
\section{Conclusion}
In this work we proposed a new strategy for microlensing
observation that not only can be used for searching MACHOs of the
galactic halo through observing the LMC stars, but also it can be
a useful tool to explore exotic space-times around compact objects
such as the NUT metric. As a result of our Monte Carlo simulation,
we obtained the detection efficiency for magnetic mass. The
minimum value for $R$ that can be observed by this method is about
$0.1$. In order to evaluate the amount of detectable magnetic mass
{\it l}, we use the relation between the magnetic mass and $R$
(Rahvar and Nouri-Zonoz 2003):
\begin{equation}
R = c\sqrt{\frac{{\it l}}{2GM}}.
\label{rl}
\end{equation}
EROS and MACHO experiments results propose that the mean value of
the mass of MACHOs is about $0.5M_{\odot}$ (Alcock et al. 2000;
Lasserre et al. 2000). It is worth to mention that this result is
obtained in the standard halo model where the mean mass of MACHOs
depends on the model that is used for the Milky Way. Assuming
standard model for the Milky Way halo, according to equation
(\ref{rl}) the minimum observable magnetic mass ${\it l}$ is
evaluated to be about $14$ m. Non-existence of magnetic mass
signal in the microlensing light curves can also put an upper
limit for the value {\it l} $<$ 14 m in the MACHOs of the Milky
Way.

\acknowledgments The authors thank M. Nouri-Zonoz, H. Hakimi
Pajouh and S. Arbabi Bidgoli for their useful comments.




\begin{figure}
\includegraphics[angle=0,scale=0.72]{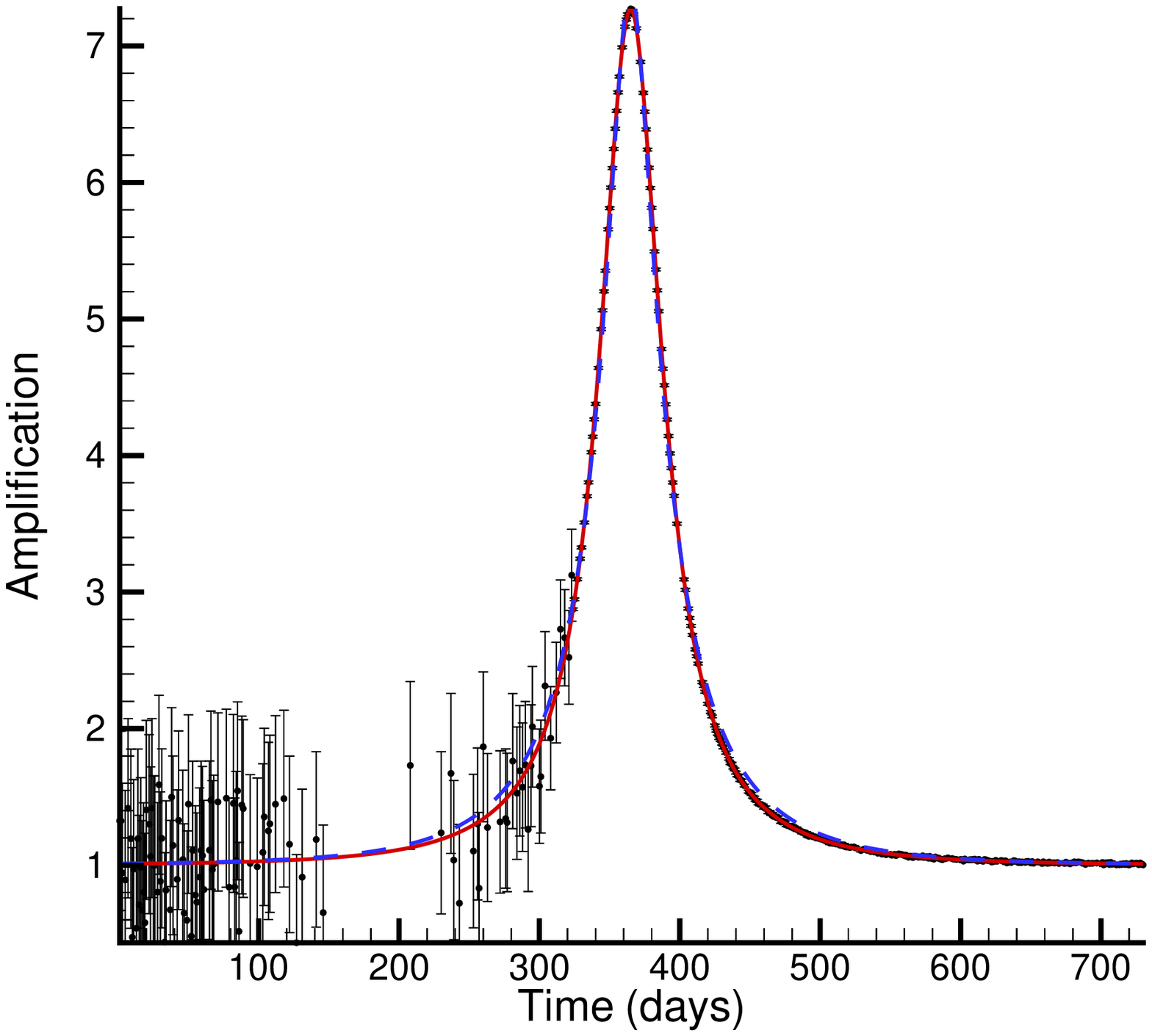}
\caption{ An example of the simulated light curves according to
our proposed observational strategy for the next generation
microlensing survey. The parameters of the light curve are chosen
to be $t_e = 100$ days, $t_0 = 365$ days, $u_0 = 0.3 $, $b=0.88$
and $R = 0.5$. The background star is chosen to have an apparent
magnitude of 22. The dashed and solid lines show the result of
least square fit of the the Schwarzschild and NUT theoretical
light curves to the simulated data, respectively. The
reconstructed NUT parameter derived from the fitting is $R_{rec}
=0.501668 $ with one sigma uncertainty of $0.001946$.
$\chi^2/N_{dof}$ for this light curve from the NUT and the
Schwarzschild fittings are 0.26 and 30.88.} \label{lc}
\end{figure}


\begin{figure}
\plotone{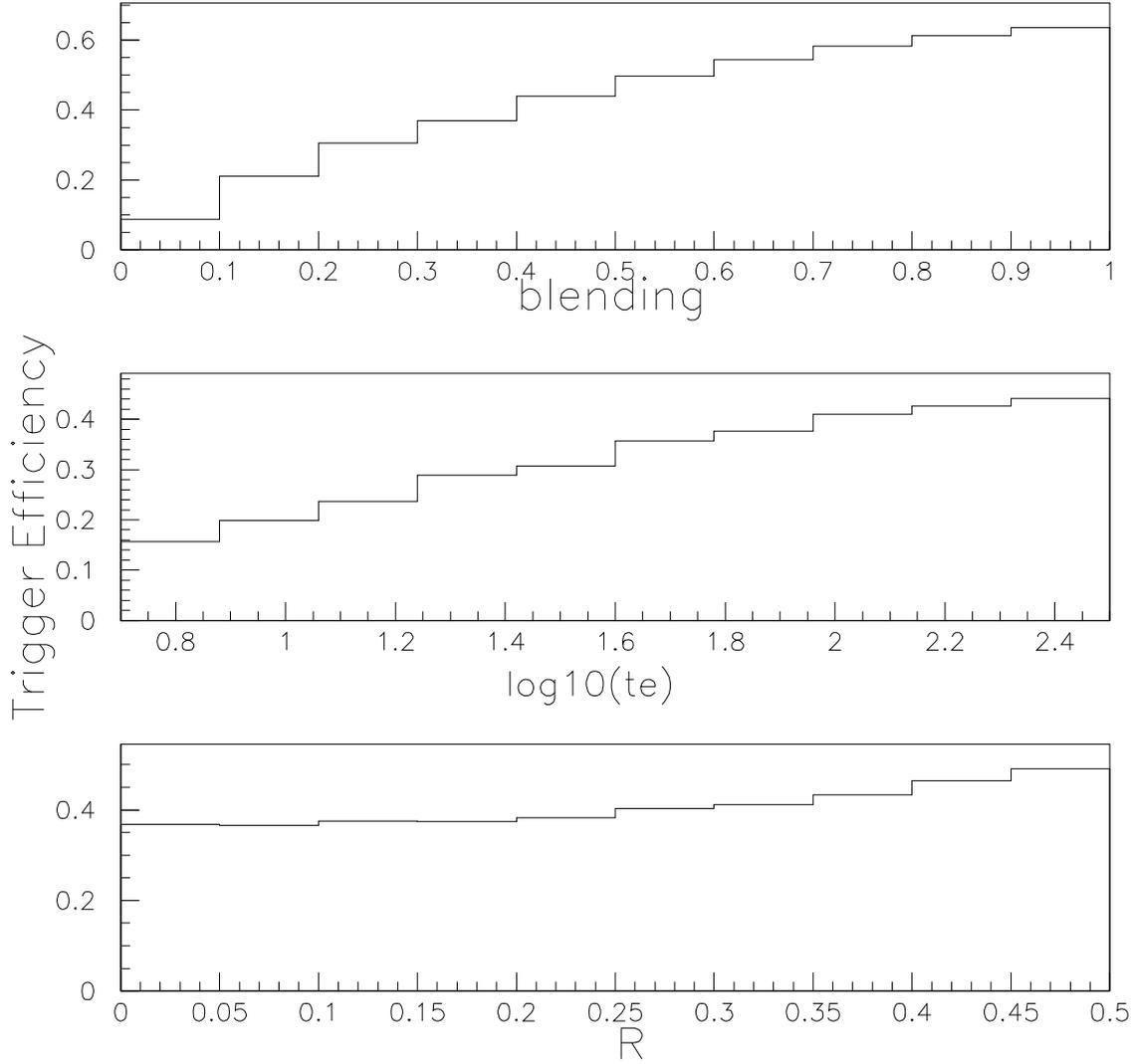} \caption{The panels from up to down show the
trigger efficiency in terms of the blending parameter, duration of
events and $R$. The efficiency of the alert system depends on the
blending parameter. This means that the bigger blending factor
produce a lower maximum magnification. Also for the long duration
events, there is a bigger chance to be alerted by the primary
telescope. For the case of events with bigger $R$s, the peak of
maximum magnification is elevated and the result is a bigger
probability for those events to be alerted.} \label{trig_proj}
\end{figure}

\begin{figure}
\plotone{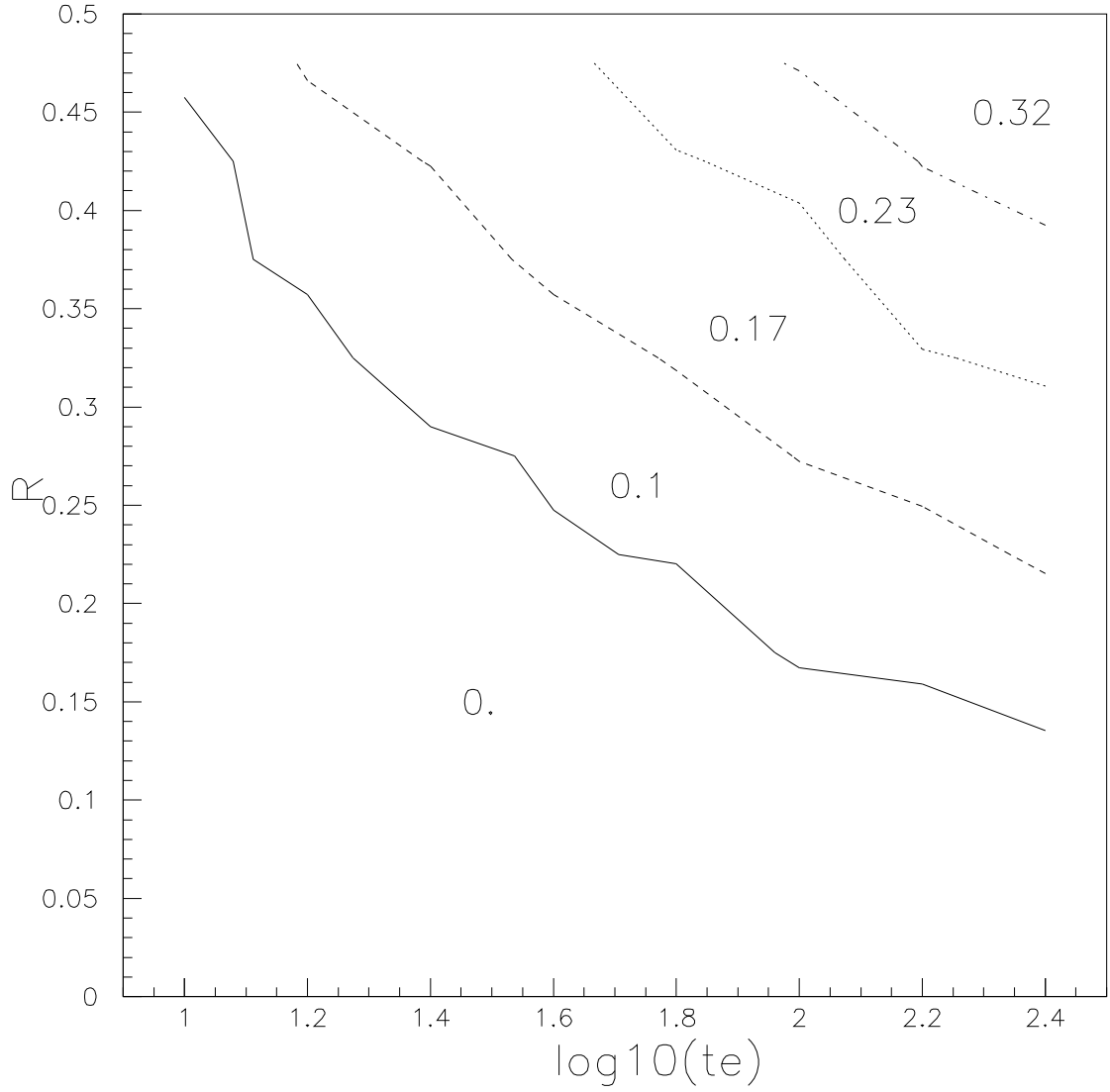} \caption{ These contours show the two dimensional
magnetic mass detection efficiency in terms of duration of events
and $R$. The numbers between the contours show the amount of
detection efficiency.} \label{2d_eff}
\end{figure}

\begin{figure}
\plotone{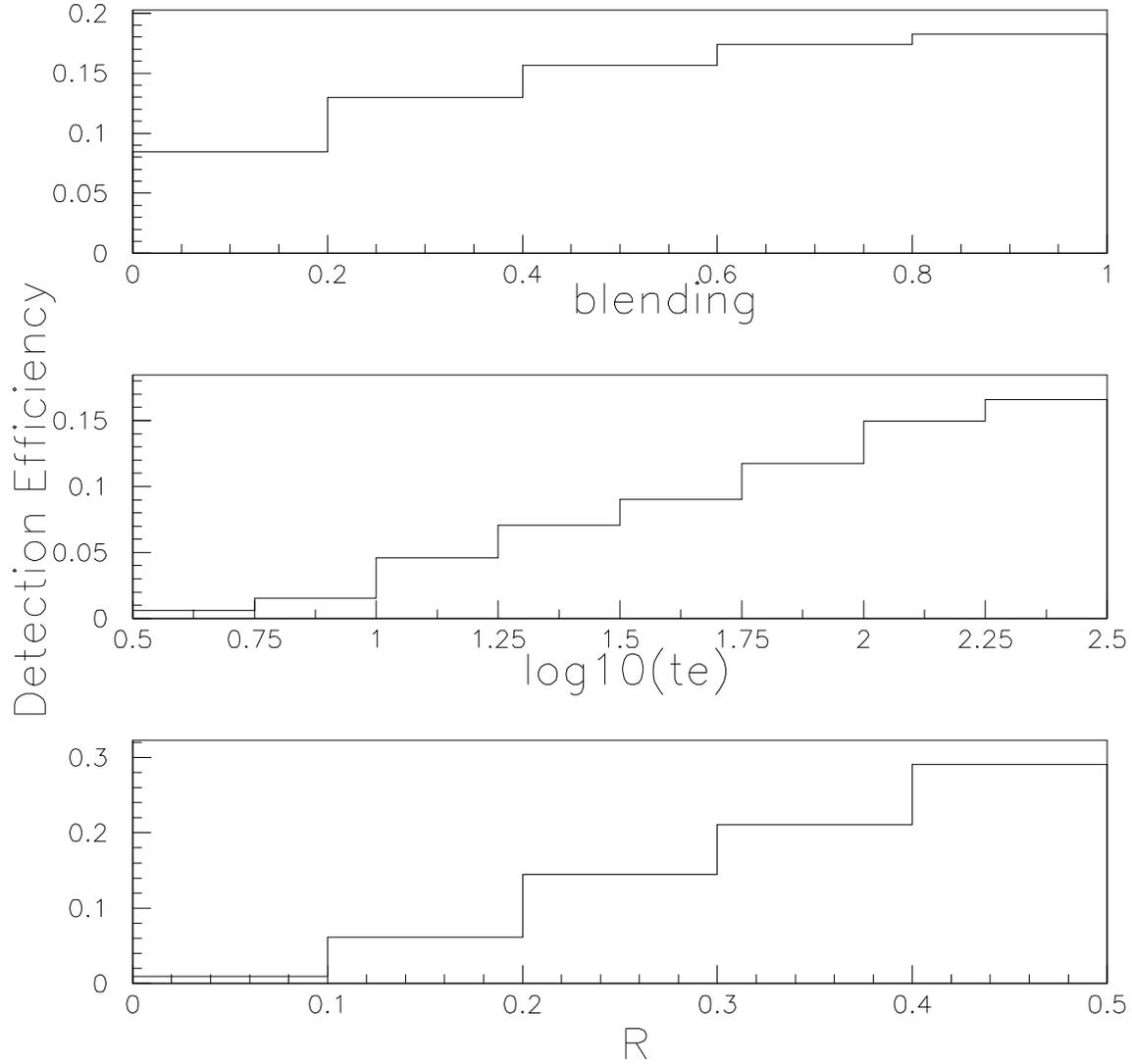} \caption{The panels from up to down show the
magnetic mass detection efficiency in terms of blending parameter,
duration of events and $R$. According to the first panel, the
magnetic mass effect can be dominated by the blending. The
detection efficiency has also direct dependence on the duration of
events and $R$. A rough value for the minimum $R$ that can be
detected is about $R=0.1$.}
\end{figure}
\label{nut_eff}
\end{document}